\documentclass{article}
\usepackage{spconf,amsmath,graphicx,hyperref}
\usepackage{amssymb}
\usepackage[table]{xcolor} 
\usepackage{booktabs}      
\usepackage{threeparttable}
\usepackage{multirow}
\usepackage{orcidlink}

\definecolor{ourmodelgreen}{HTML}{E2EFDA}
\definecolor{secondbestyellow}{HTML}{FEF2CB}

\usepackage{eso-pic}
\usepackage{ragged2e}

\newcommand{\IEEEcopyrightnotice}{%
\textcopyright\ 2026 IEEE. Personal use of this material is permitted. Permission from IEEE must be obtained for all other uses, in any current or future media, including reprinting/republishing this material for advertising or promotional purposes, creating new collective works, for resale or redistribution to servers or lists, or reuse of any copyrighted component of this work in other works.
}

\AddToShipoutPictureFG*{%
  \AtPageLowerLeft{%
    \raisebox{1.3cm}[0pt][0pt]{%
      \hspace*{\dimexpr 1in+\hoffset+\oddsidemargin\relax}%
      \parbox{\textwidth}{%
        \fontsize{7.7}{8.9}\selectfont
        \justifying
        \setlength{\parindent}{0pt}%
        \emergencystretch=1em
        \noindent
        \IEEEcopyrightnotice
      }%
    }%
  }%
}

\title{Consensus-Guided Shared-Specific Tri-View Learning for Speech Emotion Recognition}
\name{
    Bing Huang\orcidlink{0009-0000-4993-9623}$^{1\dagger}$\thanks{$^{\dagger}$ These authors contributed equally (co-first authors).}
    \qquad
    Yujian Ma\orcidlink{0009-0006-1652-5903}$^{2\dagger}$ \qquad
    Xikun Lu\orcidlink{0000-0003-0156-8805}$^{2\ast}$\thanks{${\ast}$ Corresponding author: Xikun Lu (xikunlu@stu.ecnu.edu.cn)} \qquad
    Xianquan Jiang\orcidlink{0009-0009-4360-4836}$^{3}$ \qquad
    Jinqiu Sang\orcidlink{0000-0002-4368-8787}$^{1}$
}

\address{%
  $^{1}$ School of Computer Science and Technology, East China Normal University, China\\
  $^{2}$ Shanghai Institute of Artificial Intelligence for Education, East China Normal University, China\\
  $^{3}$ Boin Hearing Technology (Shanghai) Co., LTD, China
}
\begin{document}
\ninept
\maketitle
\begin{abstract}

Speech emotion recognition (SER) benefits from heterogeneous acoustic representations, but views derived from the same utterance contain both overlapping emotional evidence and representation-dependent cues. Direct fusion may therefore propagate redundant information or obscure complementary details. To address this issue, we propose Tri-view Consensus-Guided Fusion (TriCGF) for jointly modeling spectrogram, Mel-frequency cepstral coefficients, and HuBERT representations. TriCGF organizes each view into common and view-specific components before fusion. Cross-view Consensus Learning aggregates the common components into a global reference, while View-wise Gated Integration adaptively combines this reference with each view-specific component. A soft difference regularizer further discourages excessive information overlap. Under speaker-independent evaluation, TriCGF achieves 74.19\% weighted accuracy (WA) and 75.17\% unweighted accuracy (UA) on IEMOCAP, and 94.36\% WA and 94.28\% UA on EmoDB, outperforming representative SER methods on both datasets.

\end{abstract}
%


\section{Introduction}
\label{sec:intro}

Speech emotion recognition (SER) aims to infer a speaker's emotional state from speech and supports applications in affective human-computer interaction and speech-based behavioral analysis \cite{chakhtouna2026speech, schelinski2019relation}. Emotional expression is conveyed through prosodic, spectral, and temporal variations \cite{wani2021comprehensive}, which are preserved differently across speech representations. Spectrograms retain detailed time-frequency structures \cite{badshah2017speech}, Mel-frequency cepstral coefficients (MFCCs) provide a compact description of the spectral envelope \cite{zheng2001comparison}, and self-supervised models encode higher-level contextual information \cite{hsu2021hubert,chen2022wavlm}. Since no individual representation preserves all emotion-related cues equally well, single-view SER may provide only a partial description of emotional expression.

Recent studies have strengthened emotion modeling within individual representation spaces. TIM-Net captures emotional dynamics at multiple temporal scales \cite{ye2023temporal}, while AMH-Net focuses on frequency-dependent emotional cues through formant-guided multi-band modeling \cite{li2025amhnet}. ICAPM-Net further learns relationships among acoustic parameter groups to improve both recognition and interpretability \cite{lian2026interpretable}. These methods enhance the extraction and organization of emotion-related information within a selected representation, but do not explicitly exploit the complementary cues distributed across heterogeneous views.

Multi-view SER addresses this limitation by combining representations with different acoustic properties. AMSNet integrates handcrafted and deep representations through connection attention \cite{chen2023learning}, while CA-MSER employs co-attention to model their interactions \cite{zou2022speech}. SMW\_CAT progressively fuses multiple acoustic views using cross-attention \cite{he2023multiple}, whereas Pairwise-CL promotes agreement among self-supervised, spectral, and paralinguistic representations through contrastive learning \cite{khaertdinov2024exploring}. Although these methods demonstrate the benefits of heterogeneous acoustic representations, they mainly focus on strengthening cross-view interaction or alignment. However, acoustic views derived from the same utterance are neither independent nor equally informative. Directly fusing these representations may repeatedly propagate overlapping emotional evidence while obscuring weaker but complementary view-dependent cues. The key challenge is therefore not merely how to combine multiple views, but how to organize their shared and representation-dependent information before fusion.

Shared--specific representation learning provides a useful perspective on this issue. Domain Separation Networks distinguish shared information from domain-dependent characteristics \cite{bousmalis2016domain}, while MISA and DLF separate common and modality-specific representations for multimodal affective analysis \cite{hazarika2020misa,wang2025dlf}. Disentanglement has also been explored in SER. DTNet uses identity-related information to guide the extraction of emotion-relevant features from acoustic representations \cite{10448044}. Unlike these settings, the acoustic views considered here originate from the same utterance but differ in statistical structure and abstraction level. They therefore contain overlapping emotional evidence as well as cues determined by their respective representation mechanisms, motivating explicit shared--specific organization before cross-view fusion.

\begin{figure*}[!t]
    \centering
    \includegraphics[trim={0 5 0 5}, clip, width=\textwidth]{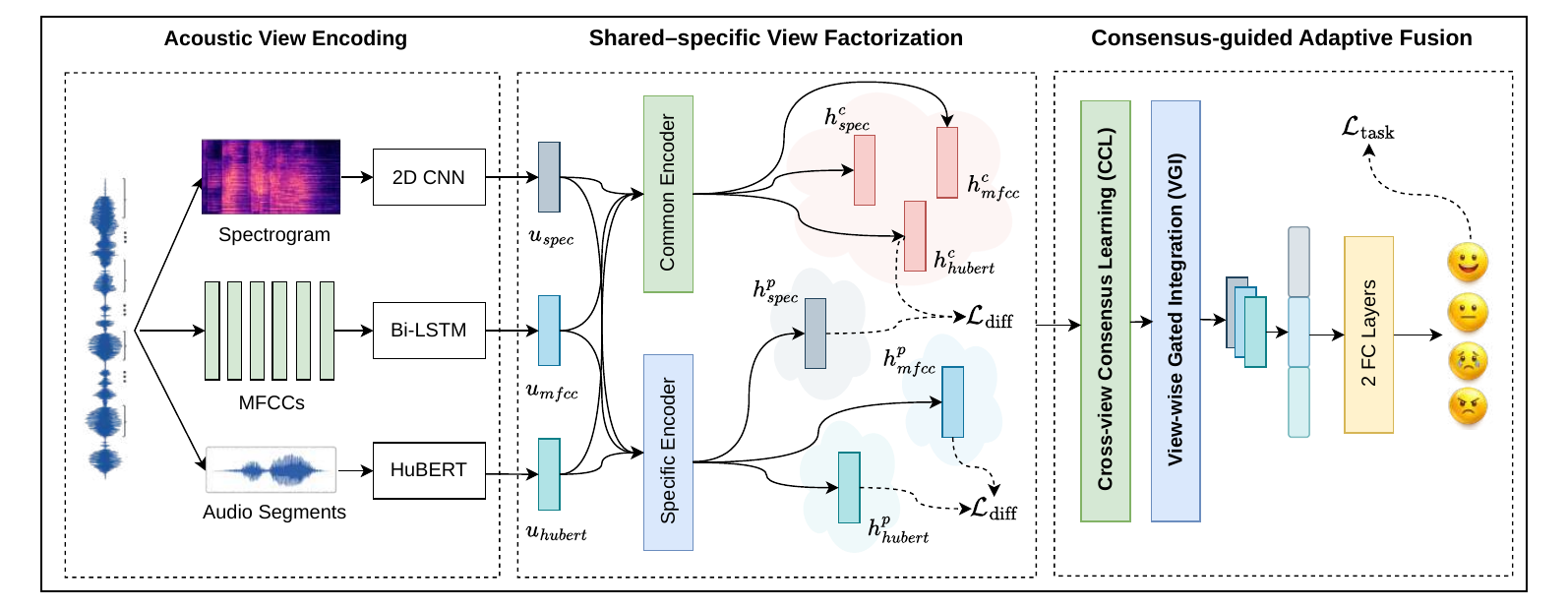}
    \caption{Overview of the proposed TriCGF framework. Three acoustic views are encoded into utterance-level representations and factorized into common ($h_m^c$) and view-specific ($h_m^p$) components, with $\mathcal{L}_{\mathrm{diff}}$ encouraging their separation. Cross-view Consensus Learning (CCL) constructs a global consensus, which is adaptively integrated with each view-specific component through View-wise Gated Integration (VGI) for emotion prediction.}
    \label{fig:tridag_arch}
\end{figure*}

To this end, we propose \textbf{Tri}-view \textbf{C}onsensus-\textbf{G}uided \textbf{F}usion (\textbf{TriCGF})\footnote{ Our source codes are available online: \url{https://github.com/Luxikun669/TriCGF}} for jointly modeling spectrogram, MFCC, and HuBERT representations. The main contributions are summarized as follows: (i) We formulate tri-view SER as a shared–specific representation problem, explicitly organizing overlapping emotional evidence and representation-dependent cues before fusion. (ii) We introduce Cross-view Consensus Learning (CCL) to form a global reference from the common representations and View-wise Gated Integration (VGI) to incorporate each view-specific component adaptively. (iii) Experiments on IEMOCAP and EmoDB demonstrate consistent effectiveness across two speaker-independent evaluation settings, while view-combination and component ablations verify the contributions of the proposed design.

\section{Proposed method}
\label{sec:format}

TriCGF integrates three heterogeneous acoustic views through shared–specific representation learning and consensus-guided fusion. As illustrated in Fig.~\ref{fig:tridag_arch}, the three views are first encoded into utterance-level representations and then factorized into common and view-specific components. The common components are aggregated into a cross-view consensus, which adaptively guides the integration of view-specific information for emotion prediction.

\subsection{Heterogeneous Acoustic View Encoding}
\label{sec:feature}

Given a speech utterance, we construct three heterogeneous acoustic views: a spectrogram, MFCCs, and the raw waveform. Their inputs are denoted by:
\begin{equation}
\mathbf{X}_{\mathrm{spec}} \in \mathbb{R}^{B \times F \times T_s},
\
\mathbf{X}_{\mathrm{mfcc}} \in \mathbb{R}^{B \times T_m \times D_m},
\
\mathbf{X}_{\mathrm{wav}} \in \mathbb{R}^{B \times L}.
\end{equation}
where $B$ denotes the batch size, $F$ is the number of spectrogram frequency bins, $T_s$ and $T_m$ denote the number of time frames, $D_m$ is the MFCC dimension, and $L$ is the waveform segment length.

The spectrogram view is encoded using a four-layer 2-D convolutional network (CNN) followed by Convolutional Block Attention Module (CBAM) \cite{woo2018cbam}, while the MFCC and waveform views are processed by a two-layer bidirectional long short-term memory (Bi-LSTM) network \cite{graves2005framewise} and HuBERT-base \cite{hsu2021hubert}, respectively. The resulting frame- or location-level features are aggregated into utterance-level representations through a learnable weighted pooling operation. Linear projections are then used where necessary to align the three representations to the same dimension:

\begin{align}
\mathbf{u}_m \in \mathbb{R}^{B \times d_h},
\quad
m \in \mathcal{M}
=
\{\mathrm{spec}, \mathrm{mfcc}, \mathrm{hubert}\}.
\end{align}
where $d_h$ denotes the unified representation dimension. During fine-tuning, only the last $k$ Transformer encoder layers of HuBERT are updated, while the remaining HuBERT parameters are frozen.
\subsection{Shared–Specific View Factorization}
\label{sec:disentangle}

The three acoustic views contain overlapping emotion-related information as well as characteristics determined by their respective representation mechanisms. To distinguish these two forms of information, each utterance-level representation is projected into a common component $\mathbf{h}_m^c$ and a view-specific component $\mathbf{h}_m^p$:

\begin{equation}
\mathbf{h}_m^c = E_c(\mathbf{u}_m; \theta^c), \quad
\mathbf{h}_m^p = E_m^p(\mathbf{u}_m; \theta_m^p).
\end{equation}

The common encoder $E_c$ is a parameter-shared two-layer multilayer perceptron (MLP). It contains a GELU activation function, dropout, and layer normalization. Applying the same encoder to all three views encourages their common components to occupy a comparable latent space. Each view-specific encoder $E_m^{p}$ adopts the same two-layer architecture but maintains an independent set of parameters. These independent mappings allow the specific branches to preserve information associated with the statistical properties and abstraction level of each representation.
\subsection{Consensus-Guided Adaptive Fusion}
\label{sec:fusion}

The factorized representations are integrated in two stages. CCL first aggregates the common representations into a global consensus. VGI then combines this consensus with each corresponding view-specific representation through a feature-wise gate, adaptively balancing common and view-specific information.

\subsubsection{Cross-view Consensus Learning} 

The three common representations are concatenated and mapped to a global consensus vector:
\begin{equation}
\mathbf{h}_{\text{global}}^c = E_{\text{gl}}\big([\mathbf{h}_{\text{spec}}^c;\mathbf{h}_{\text{mfcc}}^c;\mathbf{h}_{\text{hubert}}^c]; \theta^{\text{gl}}\big),
\end{equation}
where $[\cdot;\cdot]$ denotes feature concatenation. The consensus encoder $E_{\text{gl}}$ is implemented as a two-layer MLP that maps the concatenated 3$d_h$-dimensional representation to a $d_h$-dimensional global representation. The resulting $\mathbf{h}_{\text{global}}^c$ summarizes common emotion-related information across the three views and provides a shared reference for subsequent fusion.

\subsubsection{View-wise Gated Integration} 

For each view $m$, the global consensus $\mathbf{h}_{\text{global}}^c$ and the corresponding specific representation $\mathbf{h}_m^p$ are concatenated and passed through a view-dependent gating network:
\begin{equation}
\mathbf{g}_m = \sigma\big(G_{m}([\mathbf{h}_{\text{global}}^c;\mathbf{h}_m^p]; \theta_m^{gt})\big),
\end{equation}
where $G_{m}$ is a view-specific MLP, $\sigma(\cdot)$ denotes the sigmoid function, and $\mathbf{g}_m \in [0,1]^{d_h}$. The fused representation is computed as:
\begin{equation}
\mathbf{r}_m = \mathbf{g}_m \odot \mathbf{h}_{\text{global}}^c + (1-\mathbf{g}_m)\odot \mathbf{h}_m^p,
\end{equation}
where $\odot$ denotes element-wise multiplication. The gate adaptively balances global consensus with view-specific information. Finally, the three fused representations are concatenated for emotion prediction:
\begin{equation}
\hat{y} = \mathcal{F}_{\mathrm{cls}}\big([\mathbf{r}_{\text{spec}};\mathbf{r}_{\text{mfcc}};\mathbf{r}_{\text{hubert}}]; \theta^{cls}\big).
\end{equation}

Here, $\mathcal{F}_{\mathrm{cls}}$ denotes a two-layer MLP classifier that maps the concatenated 3$d_h$-dimensional representation to $C$ emotion classes.
\subsection{Training Objective}
\label{subsec:learning}

The overall training objective combines the emotion classification loss with a difference regularization term:
\begin{equation}
\mathcal{L} = \mathcal{L}_{\text{task}} + \alpha \mathcal{L}_{\text{diff}},
\end{equation}
where $\alpha$ controls the contribution of the regularization term.

The task loss is defined using cross-entropy:
\begin{equation}
\mathcal{L}_{\text{task}} = -\frac{1}{B} \sum_{i=1}^{B} \sum_{c=1}^{C} y_{c}^{(i)} \log \hat{y}_{c}^{(i)},
\end{equation}
where $y_c^{(i)}$ and $\hat{y}_c^{(i)}$ denote the ground-truth label and predicted probabilities of the $i$-th utterance for emotion class, respectively.

The task loss ensures that the fused representation remains discriminative for emotion recognition, but it does not directly regulate the information learned by the common and specific branches. We therefore employ a soft orthogonality-based difference loss \cite{bousmalis2016domain,hazarika2020misa}.

For a mini-batch, let $\mathbf{H}_m^c$ and $\mathbf{H}_m^p$ denote the common and specific representation matrices of view $m$. After zero-mean and $\ell_2$-norm, the difference loss is defined as:
\begin{equation}
\mathcal{L}_{\text{diff}} = \frac{1}{3}\sum_{m \in \mathcal{M}} \lVert \tilde{\mathbf{H}}_m^{c\top} \tilde{\mathbf{H}}_m^p \rVert_F^2 + \frac{1}{3}\sum_{(m_1,m_2) \in \mathcal{P}} \lVert \tilde{\mathbf{H}}_{m_1}^{p\top} \tilde{\mathbf{H}}_{m_2}^p \rVert_F^2.
\end{equation}
where $\mathcal{P} = \{(\text{spec},\text{mfcc}),(\text{spec},\text{hubert}),(\text{mfcc},\text{hubert})\}$. The first term discourages linear dependence between the common and view-specific components within each view, while the second penalizes excessive overlap among the view-specific components of different views. Both terms serve as soft regularizers rather than strict independence constraints.

\section{Experiments}
\label{sec:pagestyle}

\subsection{Datasets and Evaluation Metrics}
TriCGF is evaluated on IEMOCAP~\cite{busso2008iemocap} and EmoDB~\cite{burkhardt2005database}. 
For IEMOCAP, happy and excited are merged into a single category, resulting in four emotion classes consisting of angry, sad, happy, and neutral. The resulting subset contains 5,531 utterances and is evaluated using five-fold leave-one-session-out cross-validation. EmoDB contains 535 German utterances produced by ten speakers and is evaluated using five-fold speaker-independent cross-validation. 

We report weighted accuracy (WA) and unweighted accuracy (UA) \cite{schuller2009interspeech}. WA measures overall classification accuracy, whereas UA averages class-wise recall and is therefore less affected by class imbalance.

\begin{table}[t]
  \caption{Dataset-specific training settings for TriCGF.}
  \label{tab:hyper}
  \centering
  \footnotesize
  \begin{tabular}{lcc}
    \toprule
    \textbf{Hyperparameter} & \textbf{IEMOCAP} & \textbf{EmoDB} \\
    \midrule
    Early stop patience & 8 & 15 \\
    Batch size & 128 & 32 \\
    Initial learning rate & $1.1\times 10^{-4}$ & $8\times 10^{-5}$ \\
    Difference loss weight $\alpha$ & 0.6 & 0.5 \\
    Dropout & 0.20 & 0.30 \\
    HuBERT unfrozen layers $k$ & 6 & 8 \\
    \bottomrule
  \end{tabular}
\end{table}

\subsection{Experimental Setup}

All audio signals are resampled to 16 kHz and adjusted to a duration of 3 s. Spectrogram and MFCC features are extracted using a 40-ms Hamming window with a 10-ms hop, with 200 frequency bins retained. The model is trained using Adam with cosine learning-rate annealing for up to 100 epochs, and early stopping is applied based on validation performance. The same network architecture is used for both datasets, while the dataset-specific training hyperparameters are summarized in Table~\ref{tab:hyper}. All experiments are implemented in PyTorch and conducted on a single NVIDIA RTX 4090 GPU.



\begin{table}[t]
  \caption{Comparison with representative SER methods under five-fold speaker-independent cross-validation.}
  \label{tab:main}
  \centering
  \footnotesize
  \setlength{\tabcolsep}{3.2pt}
  \begin{tabular}{lccccc}
    \toprule
    \multirow{2}{*}{\textbf{Model}} &
    \multirow{2}{*}{\textbf{Year}} &
    \multicolumn{2}{c}{\textbf{IEMOCAP}} &
    \multicolumn{2}{c}{\textbf{EmoDB}} \\
    \cmidrule(lr){3-4} \cmidrule(lr){5-6}
    & & WA (\%) & UA (\%) & WA (\%) & UA (\%) \\
    \midrule
    SFF-NEC~\cite{thirumuru2022novel}
    & 2022 & 64.52 & 62.90 & 82.84 & 81.45 \\

    TLGCNN~\cite{yan2024speech}
    & 2024 & 66.82 & 64.21 & -- & -- \\

    WADAN+DNN~\cite{yi2020improving}
    & 2020 & 66.92 & 64.51 & 84.49 & 83.35 \\

    SkipGCNGAT~\cite{wang2024graph}
    & 2024 & 67.37 & 65.61 & -- & -- \\

    hc-former~\cite{fan2025hierarchical}
    & 2025 & 68.13 & 61.80 & 91.59 & 90.78 \\


    AMSNet~\cite{chen2023learning}
    & 2023 & 69.22 & 70.51 & 88.34 & 88.56 \\

    SAMT~\cite{shi2025speaker}
    & 2025 & 69.26 & -- & -- & -- \\

    CA-MSER~\cite{zou2022speech}
    & 2022 & 69.80 & 71.05 & -- & -- \\
    \midrule

    \textbf{TriCGF (ours)}
    & \textbf{--}
    & \textbf{74.19}
    & \textbf{75.17}
    & \textbf{94.36}
    & \textbf{94.28} \\
    \bottomrule
  \end{tabular}
\end{table}

\section{Results and Analysis}
\label{sec:results}


\subsection{Overall Results and Visualization}
Table~\ref{tab:main} compares TriCGF with representative SER methods. TriCGF achieves the best reported WA and UA among the compared methods on both datasets. On IEMOCAP, TriCGF outperforms the competitive CA-MSER method by 4.39\% in WA and 4.12\% in UA. On EmoDB, it further surpasses hc-former, yielding gains of 2.77\% in WA and 3.50\% in UA. These consistent improvements across datasets with different characteristics further demonstrate the robustness and generalization capability of the proposed framework.

\begin{figure}[!t]
    \centering
    \includegraphics[trim={65 15 65 10}, clip, width=0.49\linewidth]{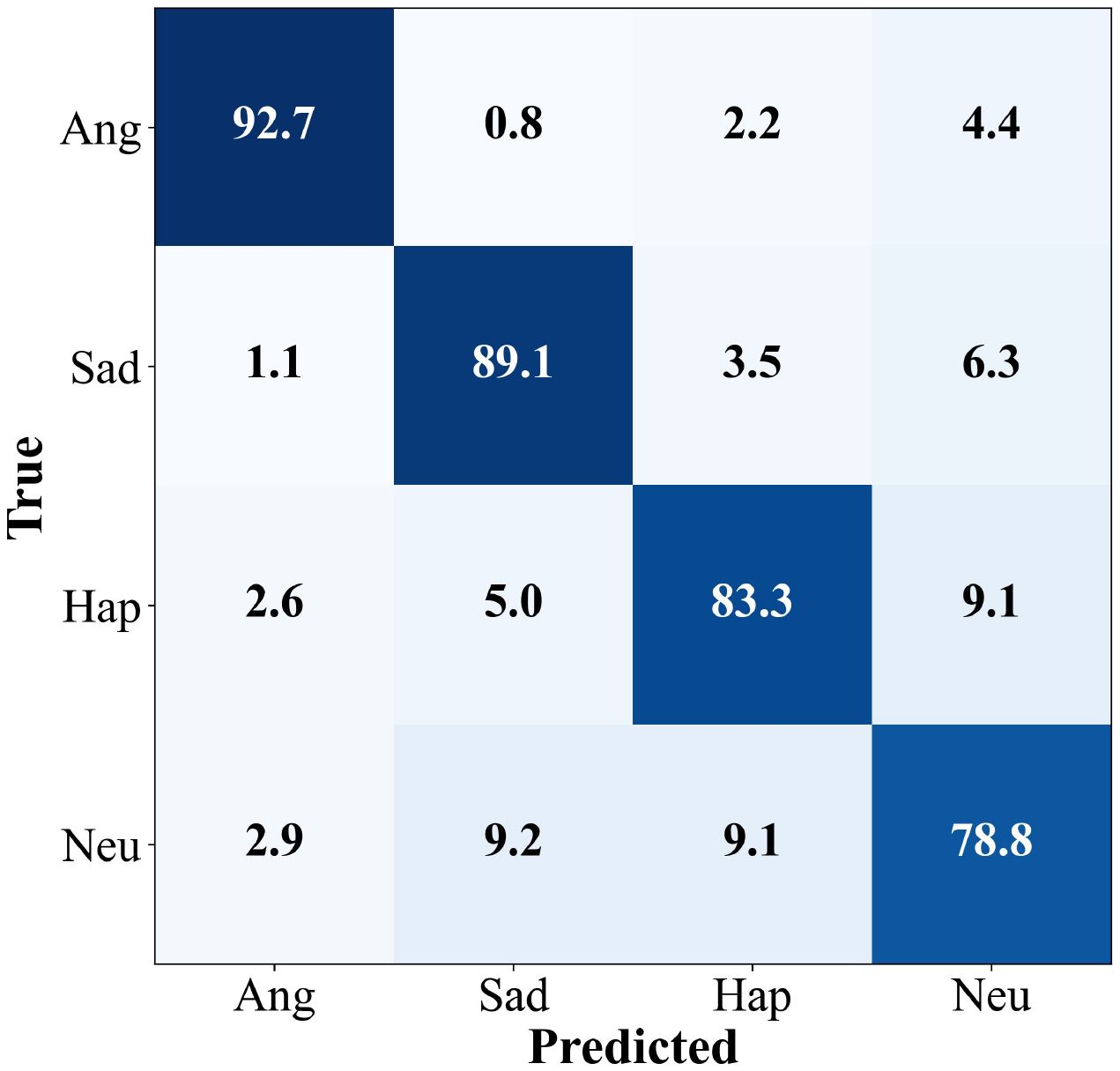}
    \hfill
    \includegraphics[trim={65 15 65 10}, clip, width=0.49\linewidth]{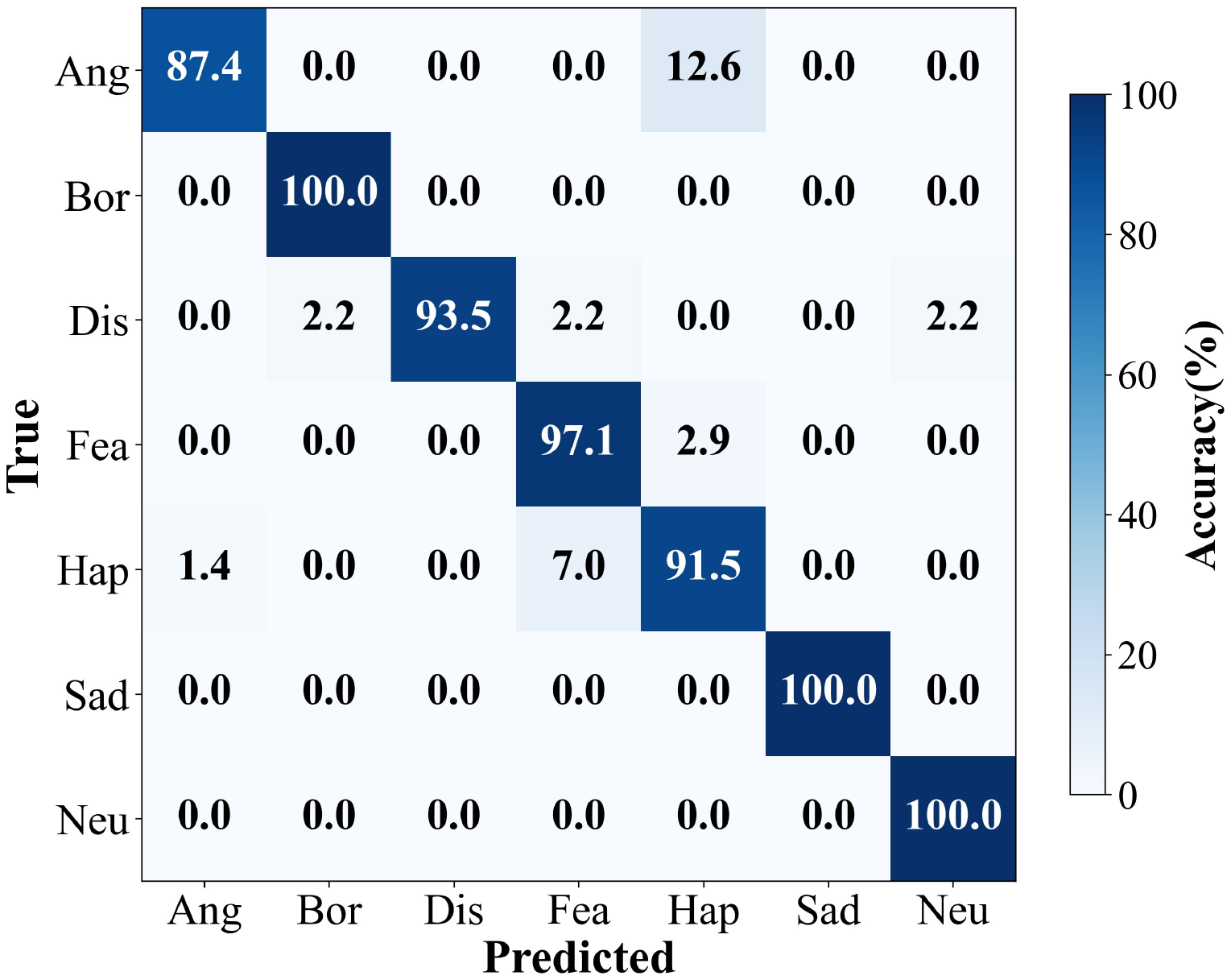}
    \caption{Normalized confusion matrices of TriCGF on IEMOCAP (left) and EmoDB (right). Rows represent ground-truth labels and columns represent predicted labels.}
    \label{fig:confusion}
\end{figure}

\begin{figure}[t]
    \centering
    \includegraphics[width=0.48\linewidth, trim=10 5 5 5, clip]{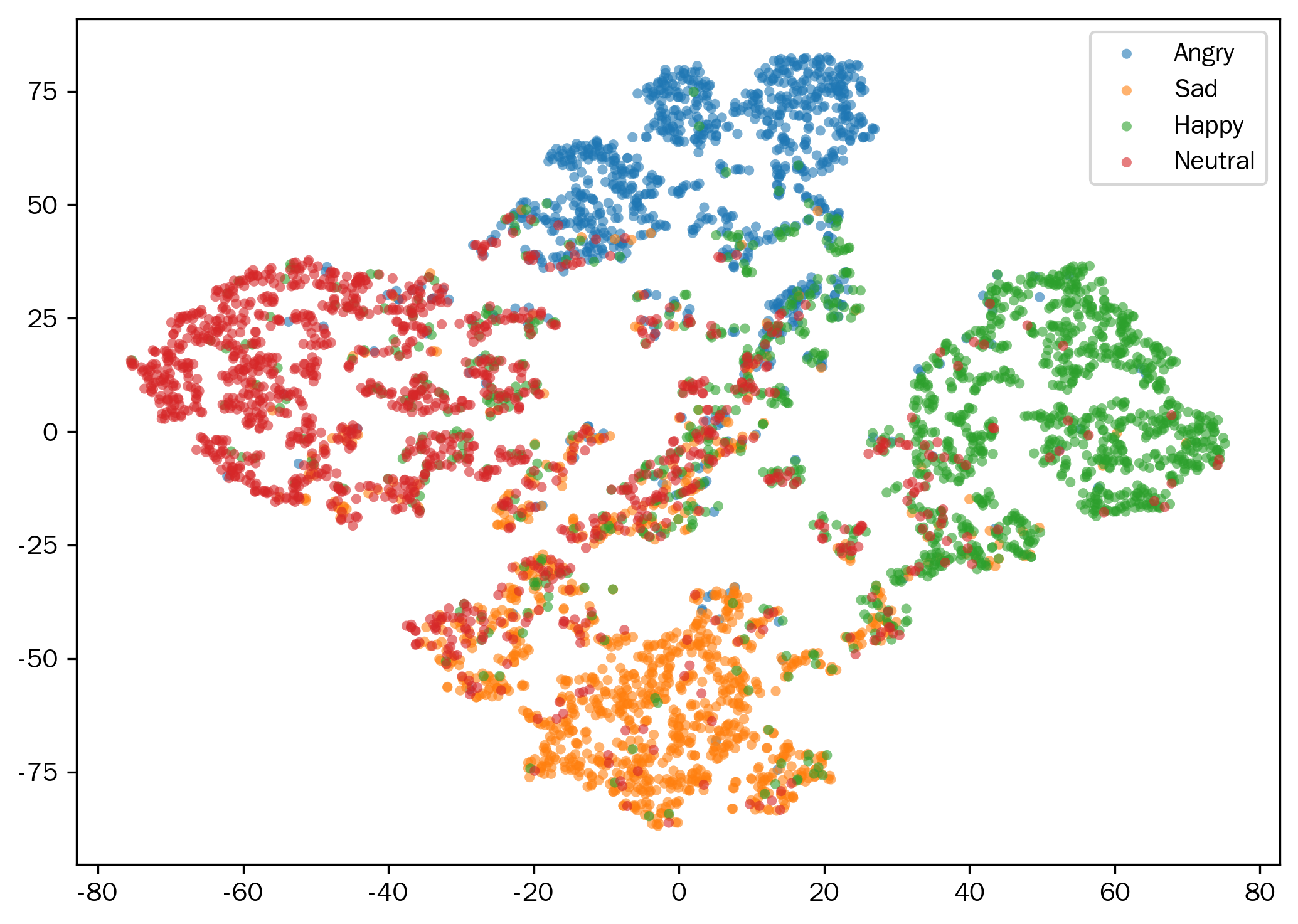}
    \hfill
    \includegraphics[width=0.48\linewidth, trim=10 5 5 5, clip]{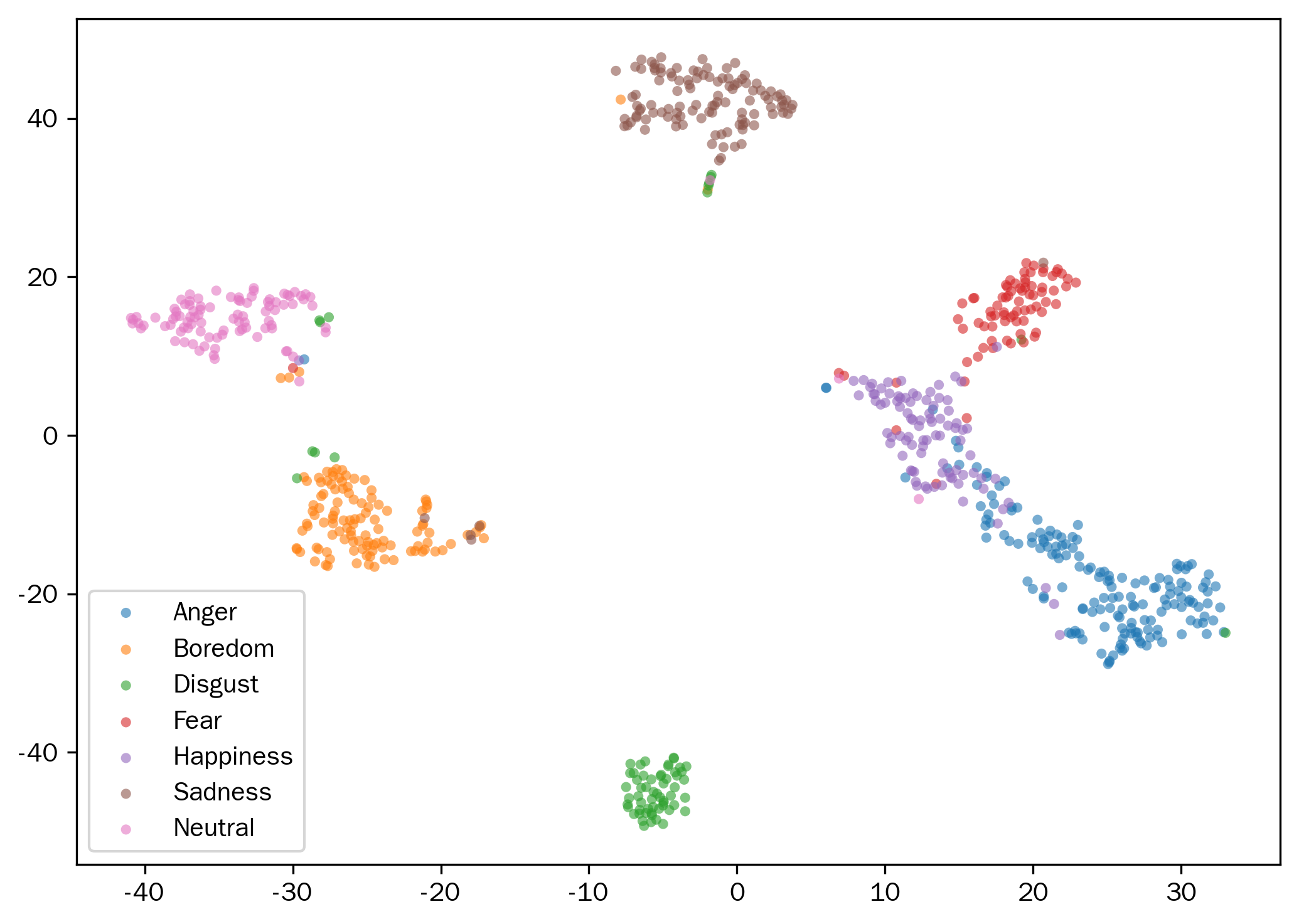}
    \caption{t-SNE visualization of TriCGF representations on IEMOCAP (left) and EmoDB (right). Colors denote emotion categories, and each point corresponds to an utterance sample.}
    \label{fig:tsne}
\end{figure}

Fig.~\ref{fig:confusion} shows the normalized confusion matrices on IEMOCAP and EmoDB. The predictions are largely concentrated along the diagonal on both datasets, which is consistent with the overall improvements in WA and UA. This indicates that TriCGF achieves stable and discriminative performance across emotion categories.

The t-SNE projections in Fig.~\ref{fig:tsne} further visualize the fused representation space. On IEMOCAP, the emotion classes form distinguishable but partially overlapping clusters, which reflects the continuous and ambiguous nature of spontaneous emotional expression. EmoDB shows more compact intra-class distributions and clearer inter-class boundaries. These patterns are consistent with the quantitative differences between the two datasets.

\begin{table}[t]
  \caption{Effect of acoustic-view combinations on TriCGF performance.}
  \label{tab:view_ablation}
  \centering
  \footnotesize
  \begin{tabular}{lcccc}
    \toprule
    \multirow{2}{*}{\textbf{View}} & 
    \multicolumn{2}{c}{\textbf{IEMOCAP}} & 
    \multicolumn{2}{c}{\textbf{EmoDB}} \\
    \cmidrule(lr){2-3} \cmidrule(lr){4-5}
     &WA (\%) & UA (\%) & WA (\%) & UA (\%) \\
    \midrule
    MFCC only & 54.69 & 56.37 & 41.51 & 35.66 \\
    Spec only & 56.43 & 57.53 & 48.33 & 45.03 \\
    HuBERT only & 72.51 & 73.70 & 90.99 & 90.01 \\
    \midrule
    Spec + MFCC & 55.95 & 57.55 & 62.24 & 56.82 \\
    MFCC + HuBERT & 72.90 & 73.79 & 90.39 & 90.26 \\
    Spec + HuBERT & 73.89 & 74.83 & 92.57 & 91.62 \\
    \midrule
    \textbf{TriCGF (tri-view)} & \textbf{74.19} & \textbf{75.17} & \textbf{94.36} & \textbf{94.28} \\
    \bottomrule
  \end{tabular}
\end{table}

\begin{table}[t]
  \caption{Ablation of the major components in TriCGF.}
  \label{tab:components_ablation}
  \centering
  \footnotesize
  \begin{tabular}{lcccc}
    \toprule
    \multirow{2}{*}{\textbf{Model Variant}} & 
    \multicolumn{2}{c}{\textbf{IEMOCAP}} & 
    \multicolumn{2}{c}{\textbf{EmoDB}} \\
    \cmidrule(lr){2-3} \cmidrule(lr){4-5}
     & WA (\%) & UA (\%) & WA (\%) & UA (\%) \\
    \midrule
    \textbf{TriCGF (full model)} & \textbf{74.19} & \textbf{75.17} & \textbf{94.36} & \textbf{94.28} \\
    w/o $\mathcal{L}_{\text{diff}}$ & 73.49 & 74.61 & 90.10 & 86.76 \\
    w/o CCL & 72.63 & 74.27 & 93.30 & 93.02 \\
    w/o VGI & 72.64 & 74.39 & 92.67 & 92.14 \\
    \bottomrule
  \end{tabular}
\end{table}

\subsection{Ablation Study}
We further analyze the contribution of different design choices through detailed ablation studies.

\subsubsection{Effect of Multi-view Design} 


Table~\ref{tab:view_ablation} reveals that the contributions of the three views are non-uniform. HuBERT provides the strongest single-view performance, reflecting the effectiveness of contextual self-supervised representations. Adding the spectrogram view consistently improves HuBERT on both datasets, suggesting that fine-grained time-frequency structures retain complementary emotional cues. The MFCC view provides a smaller and more dataset-dependent contribution when paired with HuBERT alone, but further improves the full tri-view configuration, particularly on EmoDB. These results indicate that the benefit of multi-view modeling does not arise from treating all representations as equally informative. Instead, the views provide different amounts and types of complementary information, motivating their shared-specific organization and adaptive integration.

\subsubsection{Impact of Key Components} 

Table~\ref{tab:components_ablation} reports the ablation results of the major components in TriCGF. Removing $\mathcal{L}_{\mathrm{diff}}$ causes moderate performance degradation on IEMOCAP but substantially reduces performance on EmoDB, where WA and UA decrease by 4.26\% and 7.52\%, respectively. This result indicates that the soft difference regularization is important for limiting excessive overlap between the common and view-specific representations on EmoDB. Removing either CCL or VGI also degrades performance on both datasets. On IEMOCAP, the removal of CCL and VGI produces larger decreases than removing $\mathcal{L}_{\mathrm{diff}}$, highlighting the importance of cross-view consensus construction and adaptive integration under spontaneous emotional speech. Overall, the full TriCGF achieves the best results across both datasets.

\section{Conclusion}

This work presented TriCGF, a multi-view speech emotion recognition framework that organizes common and view-specific information before fusion. TriCGF constructs a cross-view consensus from the common representations and uses it to guide the adaptive integration of view-dependent cues. Experiments on IEMOCAP and EmoDB demonstrate consistent improvements over representative SER methods, while the ablation results confirm the effectiveness of shared–specific factorization, consensus learning, and adaptive fusion. These findings indicate that the benefit of multi-view modeling depends not only on combining heterogeneous representations, but also on appropriately organizing their shared and view-specific information.




\bibliographystyle{IEEEbib}
\bibliography{refs}

\end{document}